# Social Cheesecake: An UX-driven designed interface for managing contacts

ALICIA DÍEZ
ANTONIO TAPIADOR
Departamento de Ingeniería Telemática
Universidad Politécnica de Madrid
SPAIN
adiez@dit.upm.es, atapiador@dit.upm.es

## Abstract

Social network management interfaces should consider separation of contexts and tie strength. This paper shows the design process upon building the Social Cheesecake, an interface that addresses both issues. Paper and screen prototyping were used in the design process. Paper prototype interactions helped to explore the metaphors in the domain, while screen prototype consolidated the model. The prototype was finally built using HTML5 and Javascript.

## 1 Introduction

Context collapse provokes security issues in social networks. In major social network providers, such as Facebook, users aggregate other users they know from different spheres of their life (e.g. work, family, drinking buddies, etc). The physical and temporal boundaries that exist in the offline context do not apply inside the online social network. This leads to a broad, mixed and heterogeneous audience. This phenomenon is known as context collapse [boyd 2008]. Users want to share some content in one context, e.g. the photos from last party. However, having this broad audience may lead to privacy leaks.

On the other hand, social scientists have been exploring the strength of social ties, since Granovetter [Granovetter 1977]. Inside each context, some of our contacts are strong ties. These are the people who are closest to us, which whom we spend more time interacting. Other people are weak ties, acquaintances we may only know by name. Granovetter [Granovetter 1973] suggested several categories affecting tie strength, i.e. amount of time, intimacy, intensity and reciprocal services. Beyond that, there is some recent work in models for measuring tie strength in online social networks [Gilbert 2009, Kahanda 2009, Xiang 2010]. Gilbert and Karahalios propose a model based on up to seven variables and claim to have predicted tie strength by more than the 85%. They insist in the importance of using tie strength for improving privacy controls [Gilbert 2009].

**Contact management in current social networks**

Current social networks do not take both context and tie strength aspects into account at the same time.

Facebook offers automatic lists for social contexts, like workplaces and study institutions. They also suggest lists of close friends and acquaintants. Besides, users can set up their own customized lists of friends. Nevertheless, this implies one dimension. Lists on context and tie strength cannot be easily combined. The interface is managed using plain web forms.

Google plus, launched recently, included an UX-friendly interface for contacts management, called Google Circles. This indeed takes into account the context issue, but again, tie strength is not managed at the same time. Different circles should be needed in order to group contacts by tie strength, which cannot address the two-dimensional nature of the problem.

Diaspora is another recent platform for online social networks that provides a solution for context collapse. Users can create and manage different "aspects", similar to Facebook friend-lists that are a tool for splitting contacts in several groups. As in the case of Facebook, only one dimension is used. Context and tie strength cannot be easily combined.

**Social Cheesecake**

We have designed an interface for managing contacts in an online social network that takes into account context and tie strength. A prototype model has been followed, belonging to the ambit of evolutionary development. Paper prototyping and screen prototyping have been used, across several interactions. The experience shows the limitations and advantages of both methods, as well as some lessons on the management of contacts in OSN. The prototype was built using HTML5 and Javascript.

## 2 Method

Our goal was developing an interface for contact management in online social networks that takes into account context awareness and tie strength.

This was a very open specification, so the methodology chosen was evolutionary development and prototype modeling.

The techniques used for the development of the prototype were paper prototyping and screen prototyping.

Paper prototyping [Sefelin 2003] is low-fidelity technique for the design of interfaces. Prototypes are usually cheap, demanding little time and effort expended in their made-up, so they can be through away when feedback is obtained. They are quite suitable for the innovative solution we were looking for.

Screen prototyping using Microsoft PowerPoint [Engelberg 2002] provides a prototype framework in mid-way between high fidelity and low fidelity. It allows the translation of ideas to the screen, permitting interactive content and high quality page layout, while avoiding costly time-consuming developments.

A series of quick paper prototype interactions were taken, in order to explore quickly the possible solutions for the implementation of the prototype. These interactions should be exploratory and very quick, so only one test should be made by each prototype.

Once paper prototype was squeezed, the screen prototype phase would be used. We would iterate through it until we got a stable prototype suitable for implementation.

In prototype phases, the layout was presented to the users, and a several series of questions were asked. These questions pretended to show up what comes to mind in the first place. In the case that subjects did not told us things about some interesting elements in the screen, the rest of questions tried to deepen in these aspects to learn more about them.

These are examples of questions that were presented:

- What do you think it represents?
- What can different colors mean?
- What about the circle in discontinuous path?
- What do you think about the other ellipsis in the figure? And about the number inside?
- What do you get from this diagram?
- What sense could have that button?

## 3 Results

Three different paper prototypes were needed until we could pass to the screen prototype phase.

**First interaction**

The first prototype was based on a network graph, a representation of all the closest users present in the social network. The graph was ego-centric, with the current subject represented in the center of the model. It showed the faces of contacts in the network with links between them.

It also represented clusters of people grouped by the system, each cluster belonging to one social context, as discussed in the introduction. People outside the sphere represented the people that were not added to the contact lists yet.

The goal was that users should be able to manage the contacts inside each cluster and to find more friends belonging to the same context. Search facilities should help users to find some contacts when the network had a considerable size.

Figure 1 shows the first paper prototype.

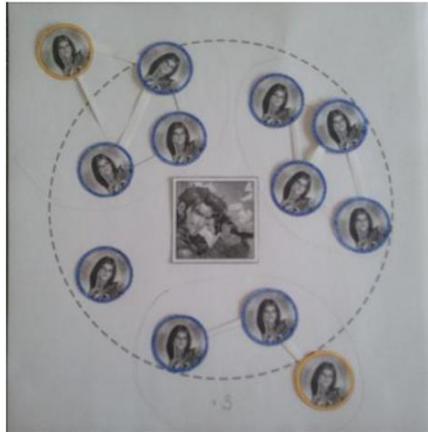
**Figure 1 Paper prototype in the first interaction**

The test showed that the layout was easily understood. Colors were not, what would probably require a legend. However, the system generated clusters were confused by the user with their own classification groups. It seemed that users did not separate these two concepts.

## Second interaction

In this interaction, two separate views were presented to users. One of them was the network graph generated by the system, with the clusters representing contexts. The other one was the user-centered sphere, with concentric spheres for strong and weak ties.

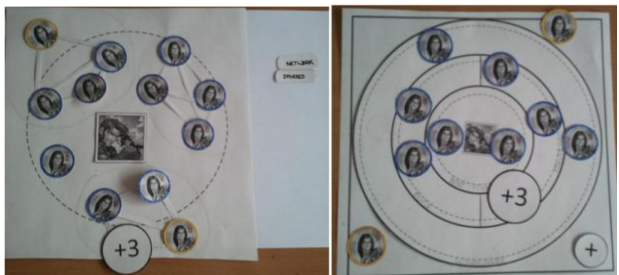
**Figure 2 Paper prototypes in the second interaction**

Results showed that the mix of the two models was too complex and confusing. The graph could be used to represent how the social network was, from an objective point of view. However, the sphere view was more useful for contact management. Sphere layers and distance to the center was something intuitive.

## Third interaction

The third interaction was only focused in the sphere model. A longitudinal sector was added, in order to add more sectors inside each layer.

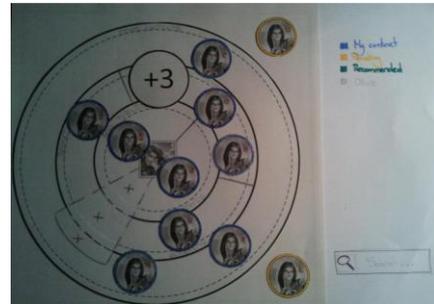
**Figure 3 Paper prototypes in the third interaction**

The test revealed that this model was too complex and had too many elements to be understood. Although it was eye-catching, the user did not know how to start using it. The grouping of contacts inside layers was not understood.

With this interaction we decided to simplify the model and pass to screen prototyping.

## Fourth interaction

For the fourth interaction, the model was simplified. The metaphor of the layout was changed from an onion-like scheme, to a cake with portions. The sphere was divided into sectors, each one having subsectors.

The prototyping method was also changed to screen prototyping, as there were several concepts already clear to us, and we also wanted to use some techniques that were not available in paper prototyping, such as animations and mouse hover.

A Microsoft PowerPoint presentation was created, consisting in 144 slides. 4 users tested the prototype, with two different profiles: 2 developers and 2 graphic designers. Tests were recorded, along with the screen, so face expressions could be registered. Tests resulted in 150 minutes of recording with its 150 minutes of cursor movement.

The test followed an outline of several tasks, including the aggregation of the first contact, the creation of a new sector when adding another contact, and contacts management.

Figure 4 shows a slide of the screen prototype, along with the video-recordings of the users.

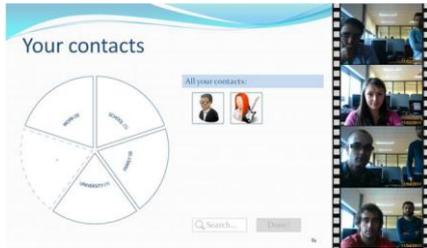

**Figure 4 Screen prototype along with video recordings of the users, in the fourth interaction**

This simpler model was clearer. Sectors were well understood by users, even when names are not created by them. There was also a clear understanding between closer and distant contacts, which represents the feature of strong and weak ties.

There were some difficulties with the interaction of the PowerPoint. Although links and hovers could be implemented, they were very limited and some of them were error prone.

However, we found few difficulties in performing the tasks, what lead us to accept this prototype as valid.

**Implementation**

The final prototype has been implemented in Javascript and HTML5 Canvas. It provides a library for managing social networks in any social networking platform.

The implementation relies on Kinetic, a Javascript library providing a base for defining canvas areas that respond differently to events (click, mouseover, mouseout..)

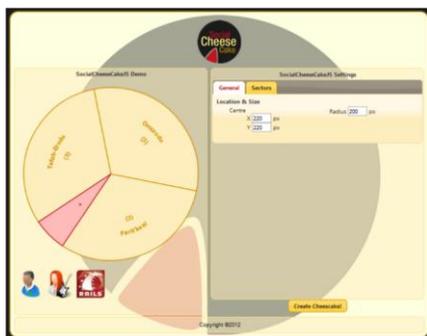

**Figure 5 Demo page with the implementation**

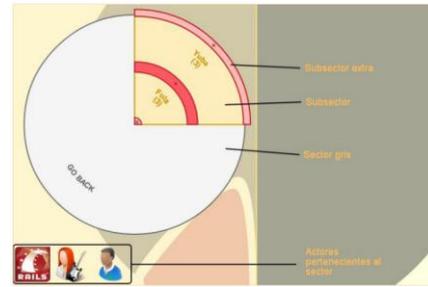

**Figure 6 Schema with subsectors in the Social Cheesecake**

The implementation is distributed as free / open source software. The source code is distributed on Github at https://github.com/ging/SocialCheesecake It includes a demo page.

## 4 Discussion

The goal of creating a new model for contact management was a very general one. In this context, paper prototyping turned out to be a very useful tool to explore metaphors in the domain of the application, contact management in this case.

We were able to iterate across quick made prototypes that did not consumed too much time. However, paper prototyping comes up with several limitations, such as the lack of animations, mouse hover, which make the prototype very less attractive model. There is neither mouse pointer, so it is difficult to give hints about drag-and-drop

Passing to screen prototype resulted in a better testing model. We were able to use animations, transitions, links and hover. Nevertheless, screen prototyping also have limitations: some elements can not be implemented, such as forms with eco-typing, dynamic search and drag & drop.

We could experience how the shape of the prototype changed considerably at the first interactions, when we where exploring the user domain. However, at the end there were less changed. This phenomenon is presented in [Buxton 2007].

The result model meets the expectations in terms of dimensionality. It provides two dimensions (sectors and subsectors) to manage context and tie strength.

Finally, acceptance tests with final users should be done in order to validate the model. In this direction, Social Cheesecake is currently being integrated as a

Facebook application, so the acceptance of this prototype can be assured.

## 5 Conclusion

Social Cheesecake is an interface for managing contacts in a social network. It is the result of a design process driven by user-experience.

Paper prototyping and screen prototyping were used to develop the model. Paper prototyping resulted very convenient when exploring the metaphors in the domain of contact management.

We could see that system generated clusters were poorly understood by users. So we needed to focus only in a ego-centric model of closest and distant spheres. This metaphor was clearly understood instead.

However, paper prototyping had some limitations as attractiveness and interaction. Screen prototyping resulted in a convenient method in the consolidation phase, but also a more costly one. Nevertheless, it still suffered from some limitations, such as drag & drop, or web forms.

The final prototype was implemented in HTML5 and Javascript, and it is freely available to be integrated in social networks platforms. Tests are being performed in order to measure final user acceptance.

## References

[boyd 2008] D.M. Boyd., Taken out of context: American teen sociality in networked publics, ProQuest, 2008.

[Buxton 2007] B. Buxton, Sketching User Experiences: Getting the Design Right and the Right Design. Morgan Kaufmann. 2007.

[Engelberg 2002] D Engelberg and A. Seffah. A framework for rapid mid-fidelity prototyping of web sites, Proc. of the IFIP 17th World Computer Congress, 2002

[Gilbert 2009] Gilbert, Eric and Karahalios, Karrie. Predicting tie strength with social media. Proceedings of the 27th international conference on Human factors in computing systems. Boston, MA, USA. 2009.

[Granovetter 1977] Granovetter, Mark S. The Strength of Weak Ties. American Journal of Sociology. The University of Chicago Press. 1973.

[Kahanda 2009] Indika Kahanda and Jennifer Neville, Using Transactional Information to Predict Link Strength in Online Social Networks. International AAAI Conference on Weblogs and Social Media. 2009.

[Sefelin 2003] Reinhard Sefelin, Manfred Tscheligi, and Verena Giller. 2003. Paper prototyping - what is it good for?: a comparison of paper- and computer-based low-fidelity prototyping. In CHI '03 extended abstracts on Human factors in computing systems (CHI EA '03). ACM, New York, NY, USA, 778-779

[Xiang 2010] Xiang, Rongjing and Neville, Jennifer and Rogati, Monica. Modeling relationship strength in online social networks. Proceedings of the 19th international conference on World wide web. Raleigh, North Carolina, USA. 2010.